\documentclass[onecolumn,prl]{revtex4}
\usepackage{graphics,graphicx}

\begin{document}

\title{Short commentary on comparing previous claim of RT superconductivity with the data of arXiv:1807.08572, "Evidence for Superconductivity at Ambient Temperature and Pressure in Nanostructures"}
\author{D. K.~Singh$^{1,*}$}
\affiliation{$^{1}$Department of Physics and Astronomy, University of Missouri, Columbia, MO 65211}
\affiliation{$^{*}$email: singhdk@missouri.edu}

\begin{abstract}
I briefly mention a previous claim of room temperature superconductivity (	arXiv:0905.3524) in Ag-based oxide material and compare their results with the most recent claim of ambient temperature superconductivity in arXiv:1807.08572. In both cases, an electrical transition to low resistance state and diamagnetism are observed. Silver is a common ingredient in both claims. Does it mean that silver holds the key to RT superconductivity or, the missing field cool data (in both reports) hint of some other physical phenomenon than superconductivity. 
\end{abstract}

\maketitle

Recent claim of ambient temperature superconductivity in Ag-Au nanostructure (arXiv:1807.08572) \cite{tr1} has created lot of excitement in scientific community. The coincidence of resistive transition with the onset of diamagnetism suggests the existence of a superconducting state in Ag-Au NS. Several years ago, Djurek et al. made a similar claim of superconducting state in Ag$_{5}$Pb$_{2}$O$_{6}$-CuO composite at 13-18$^{o}$ centigrade (286 - 291 K) (arXiv:0905.3524).\cite{tr2} In Fig. 1, we see that both systems exhibit transition to low electrical resistance states that coincide with the onset of diamagnetism. In both cases, Ag is a common ingredient. Does it mean that Ag holds the key to this most sought after physical state of material. Silver is a plasmonic material. So, it is highly surprising to find room temperature superconductivity in a system primarily consisting of Ag. 

Although the susceptibility measurement in Djurek's report was performed using ac susceptibility method, authors of the recent report (arXiv:1807.08572) seems to have followed the ZFC/FC protocol for elucidating the diamagnetic behavior. Yet, only ZFC data is shown in the arXiv report. Authors should present both ZFC and FC data for better analysis of their claim. In a superconductor, ZFC and FC curves are irreversible below the transition temperature. Yet, both curves exhibit strong diamagnetic character. On the other hand, if a system manifests some other phenomenon, then FC is typically positive. 

\begin{figure}
\centering
\includegraphics[width = 12. cm] {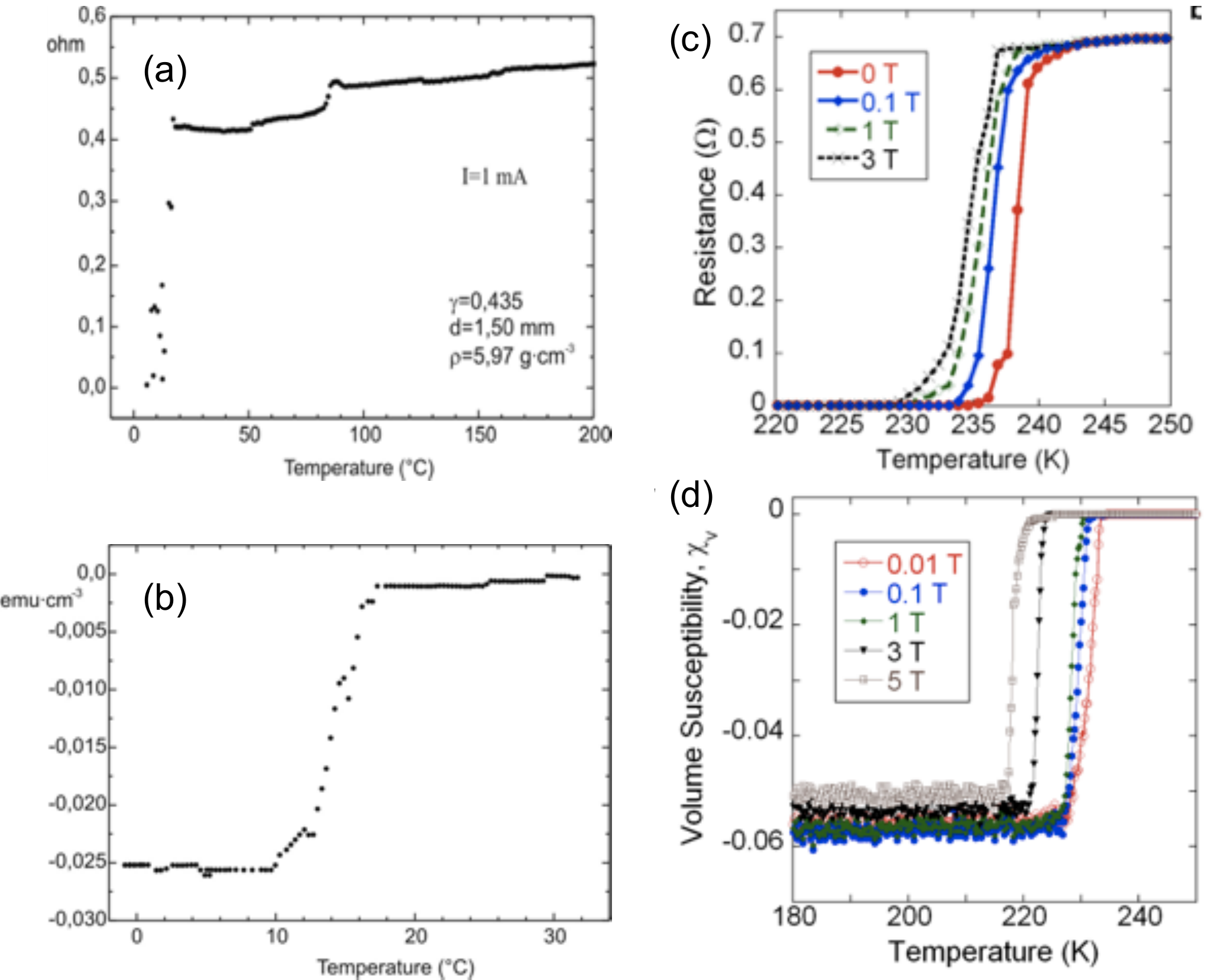} \vspace{-6mm}
\caption{ (a-b) Claimed superconducting transition in  Ag$_{5}$Pb$_{2}$O$_{6}$-CuO composite.\cite{tr2} Note that the x-axis is in centigrade. Therefore, the claimed transition at 13-18 degree is equivalent to 286 - 291 K. (c-d) Recent claim of superconducting transition in Ag-Au NS. Field cool data is missing in Fig. d, even though authors seemed to have followed the ZFC/FC measurement protocol.\cite{tr1}}
\vspace{-4mm}
\end{figure}

\clearpage

\end{document}